\documentclass[twocolumn,showpacs,preprintnumbers,amsmath,amssymb,superscriptaddress,prb]{revtex4}

\usepackage{graphicx}
\usepackage{graphics}
\usepackage{dcolumn}
\usepackage{bm}

\begin{document}

\title{Charge density wave formation in $R_{2}$Te$_{5}$ ($R$=Nd, Sm and Gd) }

\author{K. Y. Shin}
\affiliation{
Geballe Laboratory for Advanced Materials and Department of Applied Physics, Stanford University, Stanford, CA 94305 (USA)
}

\author{J. Laverock}
\affiliation{
H. H. Wills Physics Laboratory, University of Bristol, Tyndall Avenue, Bristol BS8 1TL, United Kingdom
}

\author{Y. Q. Wu}
\affiliation{
Ames Laboratory and Department of Materials Science and Engineering, Iowa State University,Ames, IA 50011 (USA)
}

\author{C. L. Condron}
\affiliation{
Stanford Synchrotron Radiation Laboratory, Stanford Linear
Accelerator Center, 2575 Sand Hill Road, Menlo Park, California 94025,
USA
}

\author{M. F. Toney}
\affiliation{
Stanford Synchrotron Radiation Laboratory, Stanford Linear
Accelerator Center, 2575 Sand Hill Road, Menlo Park, California 94025,
USA
}

\author{S.B. Dugdale}
\affiliation{
H. H. Wills Physics Laboratory, University of Bristol, Tyndall Avenue, Bristol BS8 1TL, United Kingdom
}

\author{M. J. Kramer}
\affiliation{
Ames Laboratory and Department of Materials Science and Engineering, Iowa State University,Ames, IA 50011 (USA)
}

\author{I. R. Fisher}
 \email{irfisher@stanford.edu}
\affiliation{
Geballe Laboratory for Advanced Materials and Department of Applied Physics, Stanford University, Stanford, CA 94305 (USA)
}

\date{\today}

\begin{abstract}
The rare earth ($R$) tellurides $R_2$Te$_5$ have a crystal structure intermediate between that of $R$Te$_2$ and $R$Te$_3$, consisting of alternating single and double Te planes sandwiched between $R$Te block layers.
We have successfully grown single crystals of Nd$_2$Te$_5$, Sm$_2$Te$_5$ and Gd$_2$Te$_5$ from a self flux, and describe here the first evidence for charge density wave formation in these materials.
The superlattice patterns for all three compounds are relatively complex, consisting at room temperature of at least two independent wavevectors.
Consideration of the electronic structure indicates that to a large extent these wave vectors are separately  associated with sheets of the Fermi surface which are principally derived from the single and double Te layers.
\end{abstract}


\maketitle
\section{\label{sec:level1}Introduction}

\begin{figure}[tbp]
\includegraphics[width=3.0in]{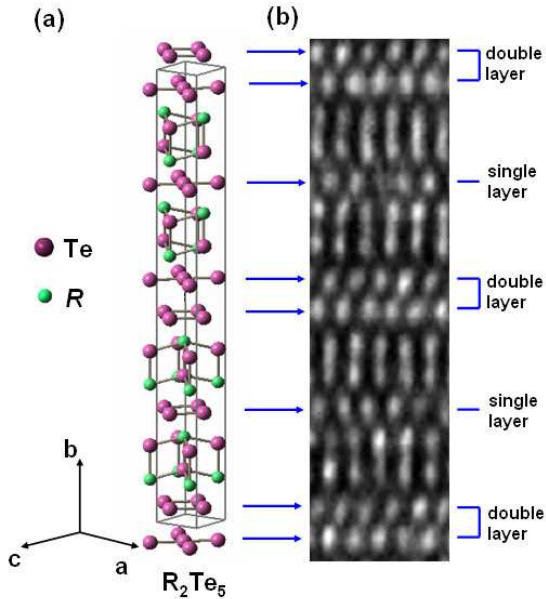}
\caption{(a)Schematic diagram showing the average (unmodulated) crystal structure of $R_2$Te$_5$.
$b$-axis is vertical in the figure. Solid grey lines indicate unit cell.
(b)High resolution TEM of Gd$_{2}$Te$_{5}$ along [101] direction, showing alternating single and double Te layers.
}
\label{fig:crystalstructure}
\end{figure}

Recently the two closely related families of compounds $R$Te$_2$ and $R$Te$_3$ ($R$=rare earth elements) have attracted considerable attention for their low dimensional electronic structure and associated charge density wave (CDW) formation.
These materials are based on single and double Te layers respectively, separated by $R$Te block layers.
Their electronic structure is especially simple, being determined by Te $p_x$ and $p_y$ orbitals in the nominally square Te planar layers.
In the case of $R$Te$_3$, large portions of the resulting quasi 2D Fermi Surface (FS) can be nested by a single incommensurate (IC) wave vector, resulting in a sharp peak in the general susceptibility $\chi(q)$.
The material suffers a CDW distortion at this wavevector for all members of the rare earth series ($R$ = La-Nd,Sm,Gd-Tm and Y), with $T_c$ values depending sensitively on $R$ due to the lanthanide contraction.\cite{Ru_2007}
In contrast, the maximum in $\chi(q)$ for the related single layer compounds $R$Te$_2$ ($R$ = La, Ce) is less well defined, and the resulting superlattice modulation varies between rare earths.\cite{Shin_2005}
The CDW gap is larger in the ditelluride than the tritelluride (for instance, the maximum gap in CeTe$_2$\cite{Shin_2005} is 600meV in contrast to 400meV for CeTe$_3$\cite{Brouet_2004} and although the CDW transition has not yet been identified in the ditelluride, transition temperatures are anticipated to be somewhat higher, too).

The title compound $R_2$Te$_5$ has an orthorhombic structure (Cmcm) as illustrated in Figure\ \ref{fig:crystalstructure}.
Note that for this space group setting, the long $b$ axis is \textit{perpendicular} to the Te planes, while the shorter $a$ and $c$ lattice parameters lie in the Te planes and are almost equal in length.
\cite{Lattice}
The material is intermediate between the two better-known families $R$Te$_2$ and $R$Te$_3$ described above, consisting of alternating single and double Te layers, separated by the same $R$Te blocks (Figure\ \ref{fig:crystalstructure}).
As we will show in this paper, the electronic structure of this material is reminiscent of the single and double layer variants, essentially comprising sheets associated with each of the Te layers separately.
The existence of this compound raises the question of whether separate modulation wave vectors might exist on the single and double Te planes separately, and if so how these wave vectors might interact or compete with each other.

Although crystals of $R_2$Te$_5$ have previously been grown from an alkali halide flux and their average structure reported, to date no superlattice modulation has been identified for this material.
In this study, we describe an alternative method to grow high quality single crystals from the binary melt, and use transmission electron microscopy (TEM) to probe the lattice modulation.
We find that all three compounds exhibit a modulation wavevector oriented along the $c^*$ axis with a magnitude close ($R$=Nd,Gd) or equal ($R$=Sm) to 2/3$c^*$, similar to that of the tritelluride compounds.
In addition, each compound exhibits at least one further set of superlattice peaks oriented away from the $c^*$ axis.
Calculations of the Lindhard susceptibility show that contributions to $\chi(\vec{q})$ enhancements arise from sections of the Fermi surface associated separately with the single and double Te planes and indicate that these different wavevectors, at least for $R$=Sm and Gd, originate from CDW formation in the double and single Te planes respectively.

\section{Crystal Growth }

High quality single crystals of $R_2$Te$_5$ ($R$=Nd,Sm,Gd) were grown by slow cooling a binary melt.
Inspection of the equilibrium binary alloy phase diagrams\cite{Massalski} reveals that $R_2$Te$_5$ has a much narrower exposed liquidus than does either $R$Te$_2$ or $R$Te$_3$, corresponding to a temperature range of less than 50$^{\circ}$C and a melt composition that varies by less than 3 at.$\%$.
Hence, for each rare earth it has been necessary to carefully determine the precise melt composition and temperature profile to achieve the optimal growth conditions that avoid the appearance of second phases.
For this reason, we have focused on just three members of the rare earth series, Nd$_2$Te$_5$, Sm$_2$Te$_5$ and Gd$_2$Te$_5$.

\begin{table}[t!]
\caption{Crystal Growth Parameters}
\label{tab:table1}

\begin{ruledtabular}
\begin{tabular}{lrr}
Crystal & Melt Composition (at.$\%$Te) & Temperature Profile \\
\hline \\
Nd$_2$Te$_5$ & 92.50$\%$  & 1050-880$^{\circ}$C \\
Sm$_2$Te$_5$ & 90.00$\%$  & 1000-920$^{\circ}$C \\
Gd$_2$Te$_5$ & 92.00$\%$  & 1050-900$^{\circ}$C \\

\end{tabular}
\end{ruledtabular}
\end{table}

Elemental starting materials of rare earth metal (Ames MPC, 99.50$\%$ for Sm and 99.80$\%$ for Nd and Gd) and tellurium(Alfa Aesar, 99.9999$\%$) were cut and placed in alumina crucibles and sealed in evacuated quartz tubes.
The ampoules were placed in a furnace and ramped to 1050$^{\circ}$C before slowly cooling to an end temperature  (Table I) at which they were removed from the furnace and the remaining flux separated from the crystals by decanting in a centrifuge.
The optimal melt composition and temperature profile varied even for the three closely spaced members of the rare earth series studied here, and are listed in Table 1.
The resulting crystals were gold in color, forming thin, malleable and micaceous plates.

Single crystal X-ray diffraction was used to confirm the phase of the crystals.
$\theta$-2$\theta$ scans along the $(0k0)$ direction revealed clear peaks for even $k$ with the appropriate lattice parameter, indicating well formed single crystalline phase of $R_2$Te$_5$.
In the measurement, it was observed that some crystals showed weak $R$Te$_3$ peaks mixed with very strong $R_2$Te$_5$ peaks.
The $R$Te$_3$ peaks could be reduced in magnitude or even caused to totally disappear by removing the surface layers of the crystals using adhesive tape.
This thin layer of $R$Te$_3$ forms on the surface of the $R_2$Te$_5$ crystals during the rapid cooling, while the remaining melt is removed by centrifuge, and is essentially a consequence of applying this growth technique to a material with such a small exposed liquidus in the phase
diagram.\cite{Massalski}

The composition of the crystals was examined by electron microprobe analysis (EMPA), for Sm$_2$Te$_5$ and Gd$_2$Te$_5$.
In both cases, tellurium content was determined to be 72 $\pm$1 at.$\%$, consistent with the value anticipated for $R_2$Te$_5$ ($5/7=71.4\%$).

\section{Transmission Electron Microscopy}

\begin{figure}[tbp]
\includegraphics[width=3.0in]{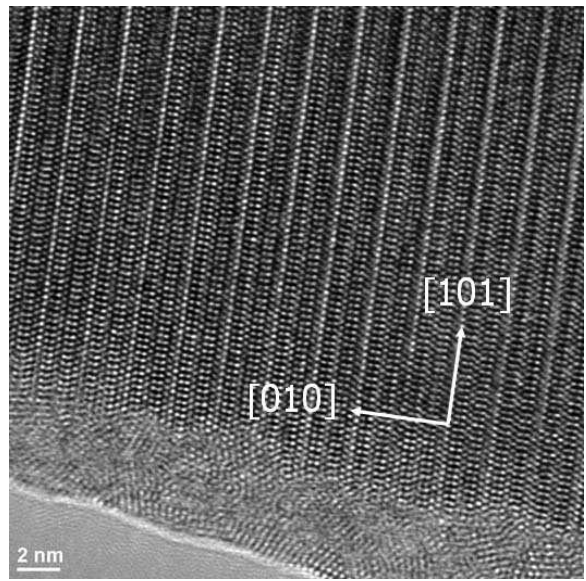}
\caption{ High resolution TEM image of Gd$_2$Te$_5$ looking down the [$\bar{1}$01]direction.
The image shows regular crystal structure over a macroscopic length scale (more than 25nm) without intercalation or stacking faults in the layering along the $b$-direction.
}
\label{fig:HRTEM}
\end{figure}

Cross-sectional TEM samples were prepared using a ``sandwich method": a small, plate-like Gd$_2$Te$_5$ single crystal ($\sim$2 x 2 x 0.1 mm) was placed in between a small stack of spacers cleaved from a single crystal silicon wafer that snuggly fit inside a 3mm diameter quartz tube.
The open spaces were in-filled using Epoxy (EPO-TEK 353ND) to hold the stack together.
A disk with a thickness about 0.5 mm was cut from the quartz tube using a diamond saw.
The Gd$_2$Te$_5$ single crystal was oriented such that the $b$-axis lay in the plane of the disk and the [101] direction was close to the disk normal.
The disk then was ground and polished to about 30$\sim$50 $\mu$m thickness.
A VCR dimpler was used to further thin the disk center area to less than 30 $\mu$m.
The final thinning of the TEM sample was performed at room temperature using a Fischione Model 1010 ion-miller.
The milling  with an initial setting of the Ar ion guns condition started at 23$^{\circ}$, 5 kV and 5 mA, at room temperature till perforation, then followed with 3 kV and 3 mA, 15$^{\circ}$ for 15 min, and final with 2 kV, 3 mA, 10$^{\circ}$ for 20 min.

A Tecnai G$^2$ F20 STEM (point-to-point resolution: 0.25 nm) operated at 200 kV was employed to do the microstructure investigation.
High resolution TEM (HRTEM) simulation was done using the National Center Electron Microscopy simulation program which employs the multi-slice approximation.

Large crystalline regions were separated by residual flux inclusions that appears continuous along the micaceaous planes.
Nonetheless, the HRTEM image of the Gd$_2$Te$_5$ single crystal shows a highly perfect crystal structure over a large area ($\sim$400 nm$^2$, Figure\ \ref{fig:HRTEM}).
Image matching to the simulations of the [101] image (Figure\ \ref{fig:crystalstructure}) suggest that the isolated bright spots are columns of Te atoms which make up the single and double layers of Te planes along the c-axis direction.
The elongated bright dumbbell features are the Gd-Te pairs in the Gd-Te block layers.

Electron beam diffraction was also measured at room temperature using a Philips CM20 FEG-TEM operating at 200kV in vacuum in order to determine the $ac$-plane modulation structure in $k$ space.
Samples were carefully cleaved to generate thin crystal pieces with thickness less then 30$\mu$m, which were mounted on a copper grid.
Optimal thickness for the measurement was achieved by making a small hole in the middle of the crystals by ion-milling in vacuum for a few hours.
Electron beams at 200kV were aligned normal to the $ac$ plane in [010] zone axis and selected area diffraction patterns (SADPs) from the flat thin edge of the crystal hole were observed at room temperature in vacuum.

\begin{figure}[t!]
\includegraphics[width=2.9 in]{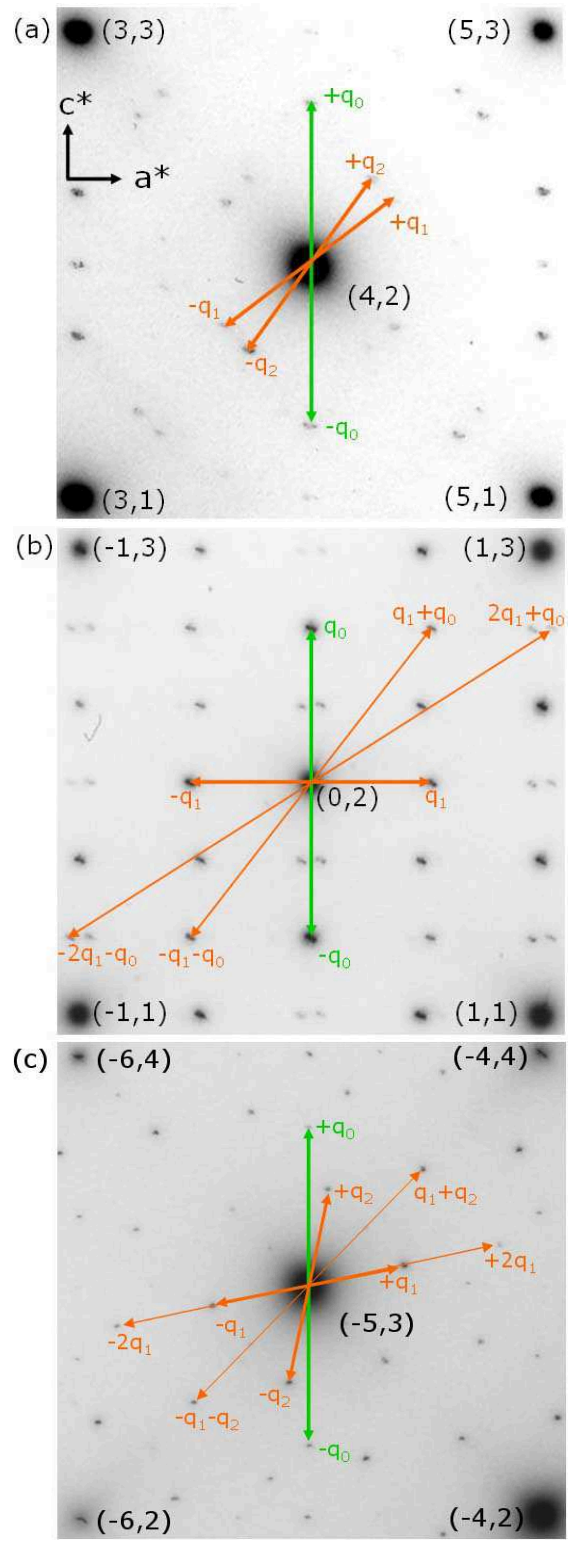}
\includegraphics[width=2.9 in]{fig3b.eps}
\includegraphics[width=2.9 in]{fig3c.eps}
\caption{SADPs along (010) zone axis for (a) Nd$_2$Te$_5$ (b) Sm$_2$Te$_5$, and (c) Gd$_2$Te$_5$.
Bragg peaks are labeled by ($h l$).
Systematic modulation wavevectors are listed in Table II.
}
\label{fig:TEM}
\end{figure}

All three compounds studied exhibit a complex set of superlattice peaks in the $ac$ plane(Figure\ \ref{fig:TEM}).
As previously observed in other families of rare earth tellurides, the satellite peaks in the quadrant defined by the Bragg peaks for $h+l=even$ translate equivalently by reciprocal lattice wave vectors $\vec{G}=(h,k,l)$, $h+l=even$.\cite{Shin_2005,DiMasi_1994,DiMasi_1995,DiMasi_1995a,DiMasi_1996}
The relative satellite peak positions in the first quadrant are listed in Table II in units of the reciprocal lattice parameters.

The SADPs for Nd$_2$Te$_5$, Sm$_2$Te$_5$ and Gd$_2$Te$_5$ are all different, but nevertheless have some common features.
In particular, all three compounds exhibit an `on-axis' superlattice reflection $\vec{q}_0$ oriented along either the a$^*$ or c$^*$ direction.
Within the resolution of TEM, we cannot distinguish $a$ and $c$ lattice parameters.
However, high resolution X-ray diffraction on Gd$_2$Te$_5$ indicates that $\vec{q}_0$ is, in fact, oriented along the c$^*$ direction.\cite{Shin_b}
For simplicity, we have also listed this lattice modulation as being along c$^*$ for Nd$_2$Te$_5$ and Sm$_2$Te$_5$ in Table II, although this remains to be confirmed.
This on-axis modulation wave vector is incommensurate for Nd$_2$Te$_5$ and Gd$_2$Te$_5$ with $\vec{q}_0$ = 0.688 $\vec{c}^*$ and 0.687 $\vec{c}^*$ for the two compounds respectively (Figure\ \ref{fig:TEM}(a),(c)).
In contrast, the on-axis wave vector for Sm$_2$Te$_5$ is commensurate within the resolution of the measurement, with $\vec{q}_0$ = 0.667$\vec{c}^*$= 2/3$\vec{c}^*$  (Figure\ \ref{fig:TEM}(b)).

\begin{table}[t!]
\caption{CDW wavevectors $\vec{q} = \protect\alpha \vec{a}^* + \protect\beta
\vec{c}^*$. }
\label{tab:table2}

\begin{ruledtabular}
\begin{tabular}{lcrr}
Crystal & Q & $\alpha$ & $\beta$ \\

\hline \\
Nd$_2$Te$_5$ & q$_0$ (on-axis)& 0.000$\pm$0.003 & 0.688$\pm$0.002 \\
 & q$_1$ (off-axis) & 0.366$\pm$0.003 & 0.269$\pm$0.003 \\
 & q$_2$ (off-axis) & 0.269$\pm$0.003 & 0.366$\pm$0.003 \\

\hline \\
Sm$_2$Te$_5$ & q$_0$ (on-axis) & 0.000$\pm$0.004 & 0.667$\pm$0.004 \\
 & q$_1$ (off-axis) & 0.521$\pm$0.004 & 0.000$\pm$0.003\\

\hline \\
Gd$_2$Te$_5$ & q$_0$ (on-axis) & 0.0000$\pm0.003$ & 0.6871$\pm0.002$ \\
 & q$_1$ (off-axis) & 0.417$\pm0.003$ & 0.083$\pm0.003$ \\
 & q$_2$ (off-axis) & 0.083$\pm0.003$ & 0.417$\pm0.003$ \\

\end{tabular}
\end{ruledtabular}
\end{table}

In addition to the on-axis CDW, each of the compounds has a distinct and unique off-axis CDW structure.

Neglecting the small difference in $a$ and $c$ lattice parameters, which is below the resolution of this measurement, the off-axis CDW in Nd$_2$Te$_5$ seems to have four fold rotational symmetry and the lattice modulation can be simply characterized by a single wavevector $\vec{q}_1$.
The second equivalent wavevector $\vec{q}_2$ is generated by reflection symmetry about the $a^*$ and $c^*$ axes.
Similar symmetry mapping has been reported for the off-axis superlattice peak $\vec{q} = 0.6 a^* + 0.2c^*$ in LaTe$_2$\cite{ Shin_2005}.

The off-axis CDW in Sm$_2$Te$_5$ is unique among the three compounds studied in that $\vec{q}_0$ and $\vec{q}_1$ generate the off-axis higher harmonics $\vec{q}_1+\vec{q}_0$ and $2\vec{q}_1+\vec{q}_0$ which are \textit{incommensurate} in the $a$-direction and \textit{commensurate} in the $c$-direction.
All the other peaks can be expressed in terms of linear combinations of q$_0$ and q$_1$ as indicated in Figure\ \ref{fig:TEM}(b), which means that the remaining other peaks are higher harmonics of those two q vectors.
Correspondingly, $\vec{q}_{0}$ = 2/3$c^*$ = 0.667$c^*$ and $\vec{q}_1 = 0.521 a^*$ are stronger in intensity than the higher harmonics $\vec{q}_1+\vec{q}_0$ and $2\vec{q}_1+\vec{q}_0$.
A similar higher harmonic CDW structure formed by linear combination of two distinct q vectors was also recently found in ErTe$_3$ by high resolution X-ray diffraction measurements.\cite{Ru_2007}

In contrast to Nd$_2$Te$_5$ and Sm$_2$Te$_5$, the off axis CDW in Gd$_2$Te$_5$ is fully commensurate in both a$^*$ and c$^*$.
High resolution X-ray diffraction measurements at Stanford Synchrotron Radiation Laboratory(SSRL)\cite{Shin_b} confirm that all peaks can be indexed by linear combinations of the two wavevectors $\vec{q}_1$=5/12 $a^*$ + 1/12 $c^*$ and $\vec{q}_2$=1/12 $a^*$ + 5/12 $c^*$, although it is not immediately clear that these are really the two fundamental wavevectors since other linear combinations are also possible in such a fully commensurate satellite peak structure.

\section{Electronic Structure}

\begin{figure}[b]
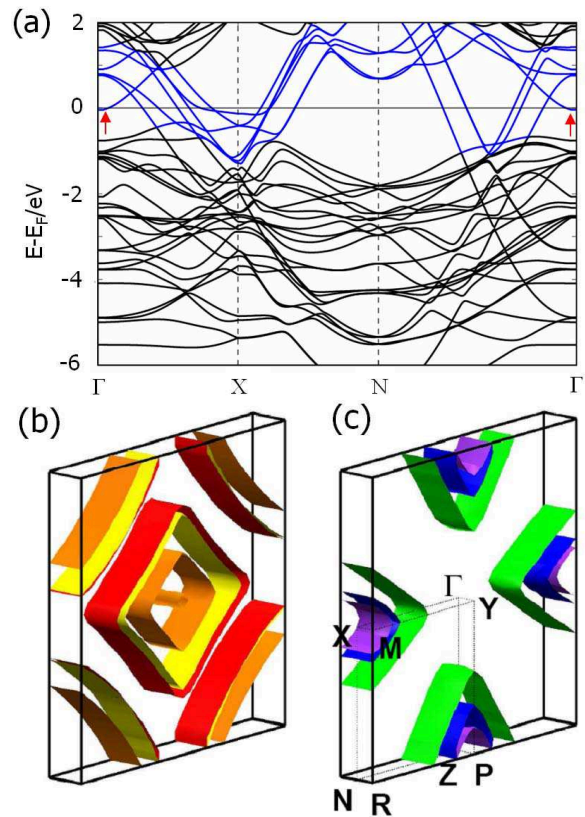

\includegraphics[width=3.0 in]{fig4a.eps}
\includegraphics[width=3.0 in]{fig4b.eps}
\caption{(Color online) (a) Band structure of $R_2$Te$_5$ calculated by LMTO for $R$=Lu. 
The blue solid lines arise from Te $5p$ states in the square planar layers.
(b) and (c) The resulting Fermi surfaces, illustrating the trilayer splitting that arises due to coupling between the three Te planes. Red and yellow sheets in (b) and blue and purple sheets in (c) indicate the weaker bilayer splitting originating from Te double layer, while orange and green sheets in (b) and (c) are due to Te single layer and the splitting is significant. 
The $\Gamma$-point at the centre of the zone and other high symmetry points have been labeled.
}
\label{fig:bandstructure}
\end{figure}

The electronic band structure for $R_2$Te$_5$ was calculated using the linear muffin-tin orbital (LMTO) method within the atomic sphere approximation including combined-correction terms as described in Refs.\onlinecite{Laverock_2004} and \onlinecite{Barbiellini_2003} and the results are shown in Figure\ \ref{fig:bandstructure} specifically for $R$=Lu (chosen to avoid the complications associated with the description of (band) $f$-electrons within the local density approximation).
The slight difference in lattice parameters originating from the structural orthorhombicity was ignored in the calculation and the value of a=c=4.36$\AA$ (b/a=10.06) from Ref.\onlinecite{DiMasi_1994} for Sm$_2$Te$_5$ was used, which is very close to a=4.34$\AA$ for SmTe$_3$.\cite{Laverock_2004}
Since the FS is comprised of Te $5p$ states originating from the Te atoms in the square planar layers, the general {\em topology} of the FS is relatively insensitive to the particular choice of rare earth atom, and indeed to changes in the lattice parameter of $\sim 5$\%, allowing us to interpret these results as prototypical for all of the other rare earth compounds.
All calculations included a basis of $s$, $p$, $d$ and $f$ states, and self-consistency was achieved at 1280 k-points in the irreducible $(1/8)^{\rm th}$ wedge of the BZ (corresponding to a mesh of $30 \times 8 \times 30$ in the full BZ).

\begin{figure}[b]
\includegraphics[width=3.0 in]{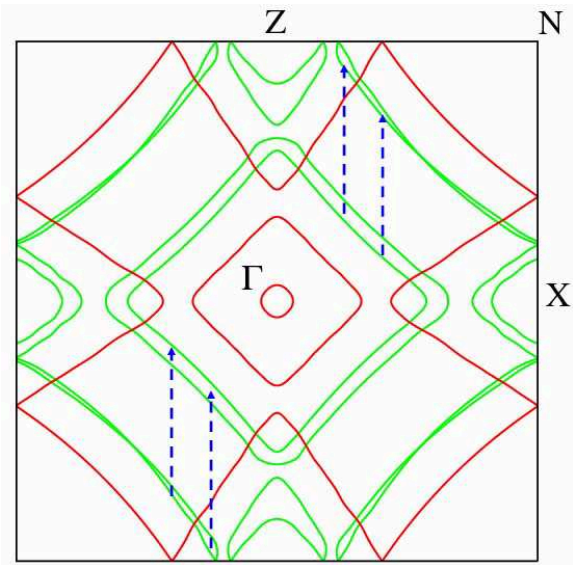}
\caption{(Color online) Fermi surface at $k_y=0$.
Red and green lines approximately indicate sections of FS, formed by
$5p$ orbitals from the single and double Te layers respectively.
Arrows indicate the on-axis lattice modulation $\vec{q}\sim2/3c^*$ observed in SADP for Nd$_2$Te$_5$, Sm$_2$Te$_5$ and Gd$_2$Te$_5$.
}
\label{fig:FSnesting}
\end{figure}

As previously found for $R$Te$_2$ and $R$Te$_3$,\cite{Laverock_2004} the electronic structure for $R_2$Te$_5$ is two dimensional and has minimal dispersion perpendicular to the Te planes(Figure\ \ref{fig:bandstructure}(b) and (c)).
Six bands formed by $5p$ orbitals from Te square planes were observed to cross the Fermi level (blue solid lines in Figure\ \ref{fig:bandstructure}(a)) and the corresponding FS at $k_y=0$ is depicted in Figure\ \ref{fig:FSnesting}.

In the tritelluride compounds, the inequivalence of the two Te atoms in the double square planar layer breaks the degeneracy of these bands, and the resulting bilayer splitting has been observed directly in ARPES studies. \cite{Gweon_1998,Brouet_2004}
For $R_2$Te$_5$, in addition to this double Te sheet, there is an additional single Te layer, and the band structure
reflects this via a triple splitting of its Te states.
The splitting between states originating from the double layer is weak and of a similar magnitude to the bilayer splitting in the tritelluride compounds, whereas the splitting between either of these double layer states and the state due to the single layer is more significant.\cite{Laverock_2004}

In addition, a small circular electron pocket due to the hybridization with a Lu \textit{d}-state above $E_F$ (indicated by red arrows in Figure\ \ref{fig:bandstructure} (a)) was observed around the $\Gamma$ point. Details of this hybridization are sensitive to the rare earth involved in the calculation, unlike the other Te $5p$ bands, and hence we might expect that the presence and exact volume of this pocket varies, as we progress through the lanthanide series.
This in turn may vary the precise location of $E_{F}$ within the Te 5$p$ bands, although small fluctuations of the electron pocket volume do not seem to significantly impact interactions between 5$p$ electrons in the Te square planes. 
In addition, the topology of this small circular section does not contribute any appreciable peak structure to the susceptibility, leading us to put less emphasis on it in the subsequent analysis of $\chi(\vec{q})$.

Even with the same $a$ and $c$ lattice parameters used in the calculation, the orthorhombicity due to the relative orientation of $R$-Te slabs in different layers is reflected in the electronic structure and produced unequal electron pockets centered at X and Z (Figure\ \ref{fig:FSnesting}).
The directional difference in the electronic structure, in turn, suggests that the band splitting at the Fermi level partially depends on the relative geometry and interactions between Te atoms in square planes and rare earth atoms in the $R$-Te slabs, even though the interplanar interaction is believed to be small.

Nevertheless, the overall topology of the FS is surprisingly similar to the individual sections of mono- and bi-layer FS structures of $R$Te$_2$ and $R$Te$_3$ (Figure 3(a) and Figure 5(a) in Ref.\onlinecite{Laverock_2004}) and thus can be approximated by superposing the corresponding Fermi surfaces of those two materials with only minor changes to account for differences in band filling.
A close investigation of the character of the wavefunction at each point supports this view and revealed that the origin of the individual FS sections can mostly be attributed to $5p$ atomic orbitals in either the ditelluride-like single Te planar layer or the tritelluride-like double Te planar layers respectively(red and green lines in Figure\ \ref{fig:FSnesting}), if the strong orbital hybridization or orbital mixing is ignored near the band crossings at the Fermi level.
This is somewhat as expected, considering that the orthorhombic structure of $R_2$Te$_5$ is intermediate of $R$Te$_2$ and $R$Te$_3$ with very close $ac$ parameters and alternating single and double Te layers along the [010] direction.

\section{Discussion}

\subsection{Electron-Phonon Coupling, Lindhard susceptibility and CDW Formation}

CDW formation is described by the second quantized Fr\"ohlich Hamiltonian,

\begin{eqnarray}
H=\sum_{n,k}\epsilon_{k}a^{\dagger}_{n,k}a^{}_{n,k}+\sum_{m,k}\hbar\omega_{q}b^{\dagger}_{m,k}b^{}_{m,k}\nonumber
 \\
+\sum_{n,m,k,q}g_{n,m,q}a^{\dagger}_{n,k+q}a^{}_{n,k}(b^{\dagger}_{m,-q}+b^{}_{m,q})
\end{eqnarray}

where $a^{\dagger}_{n,k}$ and $b^{\dagger}_{m,q}$ are the electron creation operator in the $n_{th}$ band, and phonon creation operator in the $m_{th}$ mode respectively.
The electron-phonon interaction is tuned by the coupling $g_{n,m,q}$ between the $n_{th}$ electron band and the $m_{th}$ phonon mode and the effect on the lattice distortions can be shown by obtaining the renormalized phonon mode and the dispersion relation from Equation (2):

\begin{eqnarray}
\hbar^2\ddot{Q}_{m,q}&=&-\left[\left[Q_{m,q},H\right],H\right]\nonumber\\
\ddot{Q}_{m,q}&\approx&-\omega^{2}_{m,q}Q_{m,q}-\sum_{n}g_{n,m,q}\left(\frac{2\omega_{m,q}}{M\hbar}\right)^{1/2}\rho_{n,q} \nonumber\\
\rho_{n,q}&=&-\chi_{n}(q)\sum_{m'}g_{n,m',q}\left(\frac{2\omega_{m',q}}{M\hbar}\right)^{1/2}Q_{m',q}\nonumber\\
\ddot{Q}_{m,q}&=&-\omega^{2}_{m,q}Q_{m,q}\nonumber\\
&+&\sum_{n,m'}\frac{2g_{n,m,q}g_{n,m',q}(\omega_{m,q}\omega_{m',q})^{1/2}}{M\hbar}\chi_{n}(q)Q_{m',q}\nonumber\\
\end{eqnarray}
where $\omega_{m,q}$ and $Q_{m,q}$ refer to the oscillation energy frequency and the Fourier component of the non-interacting normal coordinate of the m$_{th}$ phonon mode respectively and $\rho_{n,q}$ indicates electron density in the n$_{th}$ electron energy band.

The resultant phonon mode softening strongly depends on the strength of $g_{n,m,q}$ and $\chi_{n}(q)$.
While the coupling strength $g_{n,m,q}$ singles out the electron bands and phonon modes relevant to lattice distortions, the distortion wave vectors are selected by the peak structure in $\chi_{n}(q)$, which is mainly decided by FS nesting:

\begin{eqnarray}
\chi_{n}(\vec{q})&=&\sum_{n'\in\left\{n\right\}}^{}\chi_{nn'}(\vec{q})+\sum_{n'\notin\left\{n\right\}}^{}\chi_{nn'}(\vec{q})\nonumber\\
\chi_{nn'}(\vec{q})&=&-\frac{1}{(2\pi)^d}\int_{1BZ}^{} d\vec{k}\frac{f_{n'}(\vec{k}+\vec{q})-f_{n}(\vec{k})}{\epsilon_{n',\vec{k}+\vec{q}}-\epsilon_{n,\vec{k}}}
\end{eqnarray}
where $f_{n}(\vec{k})$ and $\epsilon_{n,\vec{k}}$ refer to Fermi-Dirac function and the energy of the electron in n$_{th}$ band.

Although it has never been easy to obtain the exact $m$,$n$ and $q$ dependence of the coupling strength $g$ theoretically or experimentally, $\chi_{n}(q)$ is relatively accessible from band structure calculations, and indeed several authors have argued the origin of CDW formation in both $R$Te$_2$ and $R$Te$_3$ in terms of simple FS nesting conditions using tight binding band calculations.\cite{Laverock_2004,Shin_2005,DiMasi_1994,DiMasi_1995,DiMasi_1995a,DiMasi_1996,Brouet_2004}
This model was successful in identifying the sections of FS which drives the CDW modulation in these compounds, and the details of the nesting was found to be dependent mainly on the \textit{topology} of the FS at the Fermi level rather than of the whole band structure.

We have used the same approach for $R_2$Te$_5$, calculating the Lindhard susceptibility $\chi(\vec{q})$ of the LMTO band structure illustrated in Figure\ \ref{fig:bandstructure} in order to examine the origin of the \textit{on-} and \textit{off-axis} CDW super lattices observed for this family of compounds. 
For computational simplicity, the two dimensional band structure at $k_y$=0 was considered for the summation in the 1st Brillouin zone(Equation 3).

However, if we assume an isotropic coupling strength $g_{n,m,q}$=$g$ and consequently calculate $\chi(\vec{q})$  by summing over all bands including inter single-double layer contributions, this quantity is found to be relatively uninformative. Broad maxima are found centered around 0.5$a^*$ and 0.5$c^*$(Figure\ \ref{fig:chiq}(a)), but otherwise there is not a well-developed peak structure that would lead one to anticipate one particular wavevector to be favored over another.

Deeper insight can be gained when more general coupling strength $g$'s, varying for individual phonon modes and Te layers, are introduced.
Motivated by the identification of distinct sections of the FS associated with the single and double Te square planes (Figure\ \ref{fig:FSnesting}), we accordingly divide the six $5p$ bands crossing the Fermi level $E_F$ into two relevant subgroups that form ditelluride-like FS sections from single Te layers, and tritelluride-like FS sections from double Te layers.
The contributions to the Lindhard susceptibility from each subgroup, $\chi_{S}(\vec{q})$ and $\chi_{D}(\vec{q})$, can be calculated as shown in Equation (4), where $S,S'$ and $D,D'$ refer to single and double layers respectively:

\begin{eqnarray}
\chi_{double}(\vec{q})=\chi_{D}(\vec{q})+\sum_{S}^{}\chi_{DS}(\vec{q})\nonumber\\
\chi_{single}(\vec{q})=\chi_{S}(\vec{q})+\sum_{D}^{}\chi_{SD}(\vec{q})\nonumber\\
\chi_{D}(\vec{q})=\sum_{DD'}^{}\chi_{DD'}(\vec{q})\nonumber\\
\chi_{S}(\vec{q})=\sum_{SS'}^{}\chi_{SS'}(\vec{q})\nonumber\\
\end{eqnarray}

Results of these calculations are shown in Figures\ \ref{fig:chiq}(b) and (c).
Inspection of these figures indicates that $\chi_{S}(\vec{q})$ and $\chi_{D}(\vec{q})$ have a more finely peaked structure than the total $\chi(\vec{q})$ (Figure\ \ref{fig:chiq}(a)).

Using this division, and following Equation (2), the resulting renormalized phonon mode dispersion is given by:

\begin{eqnarray}
\omega^{2}_{ren,m}(q)\approx\omega^{2}_{m,q}-\frac{2g_{<D>,m,q}^2\omega_{m,q}}{M\hbar}\chi_{D}(\vec{q})\nonumber\\
-\frac{2g_{<S>,m,q}^2\omega_{m,q}}{M\hbar}\chi_{S}(\vec{q})-\frac{2g_{<DS>,m,q}^2\omega_{m,q}}{M\hbar}\sum_{DS}\chi_{DS}(\vec{q})\nonumber\\
\end{eqnarray}

This dispersion relation (Equation 5) explicitly shows how the phonon mode softening depends on the average coupling strengths $g_{<S>}$, $g_{<D>}$ and $g_{<SD>}$, and the Lindhard susceptibility contributions $\chi_{S}(q)$, $\chi_{D}(q)$ and $\chi_{SD}(q)$, from the single, double and inter single-double layer contributions respectively. This division makes it possible, at least in principle, to identify the relative coupling strength as well as the most relevant electron bands from the observed lattice distortions.
In the following two sections, we address the origin of the \textit{on}- and \textit{off-axis} lattice modulation with reference to these contributions.

\subsection{Origin of the \textit{On-axis} Lattice Modulation}

\begin{figure}[t!]
\includegraphics[width=2.90 in]{fig6a.eps}
\includegraphics[width=2.90 in]{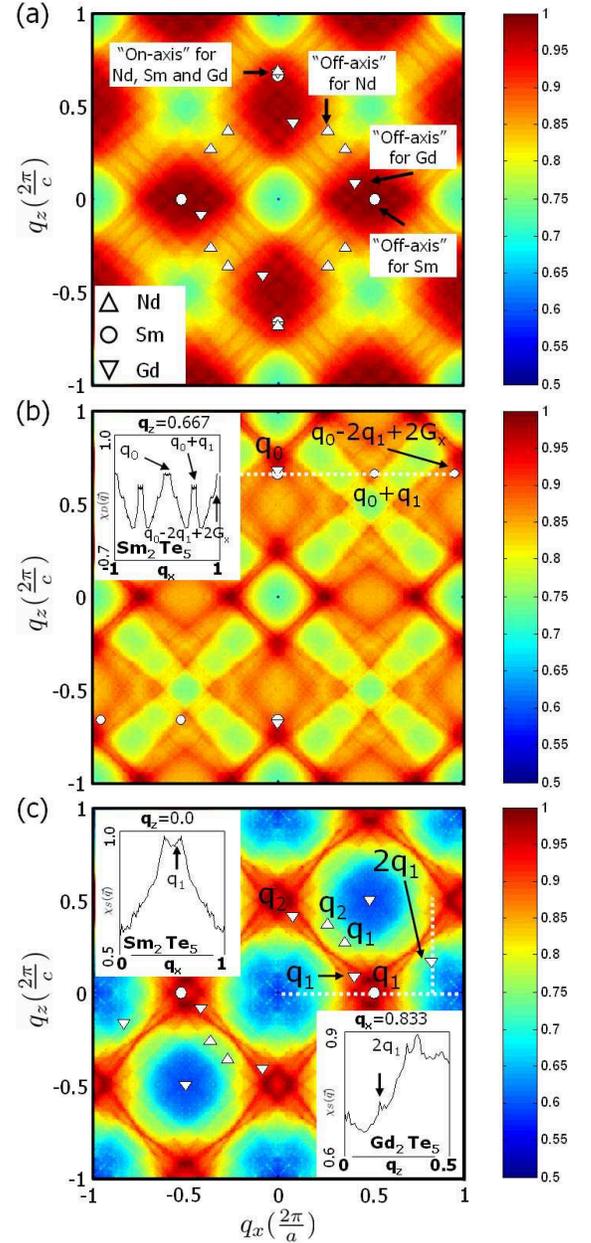}
\includegraphics[width=2.90 in]{fig6c.eps}
\caption{(Color online) Color scale for all panels: red high, blue low. 
(a) The Lindhard susceptibility $\chi(\vec{q})$=$\sum_{nn'}\chi_{nn'}(\vec{q})$ at $q_y$=0, summed for all the bands including inter single-double layer pairs. 
(b) $\chi_{D}(\vec{q})$: contribution to $\chi(\vec{q})$ from the double Te layers. \textit{On-axis} wave vectors for all three compounds lie on the \textit{global maximum}.
Inset: a line cut following the dashed line, illustrating the resonant enhancement of the mixed harmonics $\vec{q}_0+\vec{q}_1$ and $\vec{q}_0-2\vec{q}_1+2\vec{G}_x$ for Sm$_2$Te$_5$.
(c) $\chi_{S}(\vec{q})$: contribution to $\chi(\vec{q})$ from the single Te layers.
Symbols represent first harmonics of the \textit{off-axis} modulation vectors for all three compounds, and second harmonics for Gd$_2$Te$_5$.
Upper inset: a line cut following the horizontal dashed line. The \textit{off-axis} modulation lies close to the \textit{global maximum} of $\chi_{S}(\vec{q})$ for Sm$_2$Te$_5$.
Lower inset: a line cut following the vertical dashed line, showing the resonant enhancement of the commensurate \textit{off-axis} lattice modulation for Gd$_2$Te$_5$. 
}
\label{fig:chiq}
\end{figure}

All three compounds exhibit an on-axis lattice modulation with wave vector oriented along either the a* or c* axis. As noted in section III, TEM cannot distinguish these two lattice parameters, but high resolution x-ray diffraction experiments for Gd$_2$Te$_5$ have determined that $q_0$ is in fact oriented along the c* direction for this compound.
The orientation of the on-axis wave vector for Nd$_2$Te$_5$ and Sm$_2$Te$_5$ remains to be determined, but for simplicity we have referred to these as also lying along the c* direction.
The magnitude of the on-axis wave vectors for the three compounds are very similar, being commensurate $q_0=2/3=0.667$ c* for Sm$_2$Te$_5$, with very close incommensurate values for Nd$_2$Te$_5$ and Gd$_2$Te$_5$ (Table II).

Significantly, the on-axis wave vectors for all three compounds correspond to a very well-defined sharp maximum in  $\chi_{D}(\vec{q})$ calculated from the sections of the FS associated with the double Te layers (Figure\ \ref{fig:chiq}(b)).
A substantial fraction of the FS coming from these double layers are nested by this wave vector, as indicated by vertical arrows in Figure\ \ref{fig:FSnesting}.
The actual modulation wave vectors are very close to the maximum in $\chi_{D}(\vec{q})$, and $\chi_{D}(\vec{q}_{0})$ is smaller by less than 5$\%$ compared to the calculated \textit{global} maximum.
This striking correspondence is highly suggestive that the double Te layers drive the \textit{on-axis} CDW.

The simpler double-layer compound $R$Te$_3$ also exhibits an on-axis superlattice modulation, also corresponding to a similarly well-defined peak in the susceptibility.
In that case, the wave vector $q_0$$\sim$5/7$c^*$=0.71$c^*$ over the entire range of the compounds($R$=La-Tm),\cite{Ru_2007} which is also very close to $q_0$ for $R_2$Te$_5$.
The difference in $q_{0}$ values between the two families of compounds can be attributed to differences in band filling, as well as to the more complicated electronic structure in $R_2$Te$_5$.
Nevertheless, the similarity in the nesting mechanism driving the \textit{on-axis} modulation in the two compounds, and its stability across the rare earth series for both families, is remarkable.

\subsection{Origin of the \textit{Off-axis} Lattice Modulation}

Given the clear correlation between the on-axis modulation wave vectors and $\chi_{D}(\vec{q})$, it is natural to reason that the off-axis wave vectors might be more closely associated with the single Te planes.
Indeed, with the possible exception of Nd$_2$Te$_5$, none of the three compounds studied exhibit any obvious correlation between the off-axis wave vectors and $\chi_{D}(\vec{q})$, whereas, as we show below, there is a close correlation with $\chi_{S}(\vec{q})$, at least for Sm$_2$Te$_5$ and Gd$_2$Te$_5$.
However, the off-axis modulation is not as simple to account for as the on-axis modulation.
In particular, there is considerable variation in the off-axis wave vectors for the three compounds studied (Table II), reminiscent of the variation in the lattice modulation for the simpler single-layer compound $R$Te$_2$ ($R$ = La - Er).
Consequently, we consider each of the three compounds separately below.

First, Sm$_2$Te$_5$. For this compound, the off-axis modulation lies along the $a$-axis, with an incommensurate wave vector $\vec{q}_1=0.521a^*\approx0.5a^*$ (here we preserve the notation ``\textit{off-axis}" to indicate that $\vec{q}_1$ is not oriented parallel to $\vec{q}_0$, even though, in this case, $\vec{q}_1$ lies along a high symmetry direction).
As can be seen from the inset to Figure\ \ref{fig:chiq}(c), this wave vector lies very near to the \textsl{global} maximum in $\chi_{S}(\vec{q})$, indicating that the single Te layers do indeed play a significant role in driving this off-axis CDW. The behavior is also reminiscent of the unit cell doubling associated with the CDW superlattice found in some of the rare earth ditellurides $R$Te$_2$.\cite{Shin_2005,DiMasi_1996}

As described in section III, Sm$_2$Te$_5$ is unique among the three compounds studied in that the mixed harmonics of the CDW modulations $\vec{q}_0\pm\vec{q}_1$ and $\vec{q}_0\pm2\vec{q}_1$ are evident in SADP patterns.
Close inspection of the inset to Figure\ \ref{fig:chiq}(b) reveals that these wave vectors, which have a different incommensurate/commensurate structure in the $a$ and $c$ directions, are in fact closely associated with noticeably significant peak structures in $\chi_{D}(\vec{q})$.
The same figure also shows that $\vec{q}_0-2\vec{q}_1+2\vec{G}_{(100)}$, equivalent to $\vec{q}_0-2\vec{q}_1$, is actually very near to another global maximum in $\chi_{D}(\vec{q})$ adjacent to (101).
This maximum is equivalent to that which is close to $\vec{q}_0$ by a simple reciprocal lattice translation, suggestive of a resonance in the interaction due to the crystal symmetry:

\begin{eqnarray}
\frac{2g^{2}_{<D>,m,\vec{q}=\vec{q}_0-2\vec{q}_1}\omega_{m,\vec{q}=\vec{q}_0-2\vec{q}_1}}{M\hbar}\chi_{D}(\vec{q}_0-2\vec{q}_1)\nonumber \\
=\frac{2g^{2}_{<D>,m,\vec{q}=\vec{q}_0-2\vec{q}_1}\omega_{m,\vec{q}=\vec{q}_0-2\vec{q}_1}}{M\hbar}\chi_{D}(\vec{q}_0-2\vec{q}_1+2\vec{G}_{(100)}),\\
\chi_{D}(\vec{q}_0-2\vec{q}_1+2\vec{G}_{(100)})\approx\chi_{D}(\vec{q}_0)\nonumber
\end{eqnarray}

These observations suggest a significant coupling between the two wave vectors $\vec{q}_0$ and $\vec{q}_1$ in Sm$_2$Te$_5$.
Even though the off-axis modulation is principally driven by the single Te-planes (i.e. $\vec{q}_1$ is very close to the global maximum in $\chi_{S}(\vec{q})$) nevertheless, it is not insensitive to the double layers.
In contrast, the on- and off-axis CDW modulations in Nd$_2$Te$_5$ and Gd$_2$Te$_5$ seem to be independent or minimally coupled to each other, without any commensurate/incommensurate mixing.

In contrast to Sm$_2$Te$_5$, the off-axis CDW in Gd$_2$Te$_5$ is fully commensurate.
Although the modulation wave vectors $\vec{q}_{1}$ and $\vec{q}_{2}$ for Gd$_2$Te$_5$ are different to that observed in Sm$_2$Te$_5$, both lie close to global maxima in $\chi_{S}(\vec{q})$, suggesting that the single Te planes play the dominant role in driving the off-axis CDW for this compound, too.
It is worth noting, however, that the peak structure in $\chi_{S}(\vec{q})$ (Figure\ \ref{fig:chiq}(c)) is far less well developed than that in $\chi_{D}(\vec{q})$ (Figure\ \ref{fig:chiq}(b)).
Rather than a single \textit{global} maximum with little in the way of additional features, $\chi_{S}(\vec{q})$ exhibits a broad range of maxima along sharp ``ridges" (dark red regions in Figure\ \ref{fig:chiq}(c)) connecting four relatively sharp local peaks centered close to $\vec{q}$=(0 0 0.5),(0.5 0 0),(-0.5 0 0) and (0 0 -0.5).
The resulting figure is reminiscent of similar calculations for the simpler single layer compound  $R$Te$_2$, which also lack a well-defined single peak, and for which the superlattice modulation vectors also vary across the rare earth series.\cite{Shin_2005}
We will return to the variation in the off axis wave vectors later.

The two commensurate CDW vectors, $\vec{q}_{1}$ and $\vec{q}_{2}$, in Gd$_2$Te$_5$ span the entire commensurate CDW super lattice peaks in $\vec{k}$ space.
Such an extensive commensurate modulation structure may not allow a simple explanation in terms of perturbative higher harmonics.
However, it is interesting to note that $2\vec{q}_{1}$, a second harmonic of $\vec{q}_{1}$, lies exactly on top of an additional weak local maximum in $\chi_{S}(\vec{q})$, which may give some hint as to the origin of this extensive commensurate structure (inset to Figure\ \ref{fig:chiq}(c)-the strength of this feature depends sensitively on details of the calculation, but appears to be robust).
Specifically, rather than just a simple perturbation of $\vec{q}_{1}$, 2$\vec{q}_{1}$ itself also seems to be directly coupled to the relevant electronic structure through the \textit{local maximum peak} at
 $\chi_{S}(2\vec{q}_{1})$.
This effect coherently enhances the CDW instabilities at $\vec{q}$=2$\vec{q}_{1}$ and seems to help the commensuration mechanism of the off-axis CDW to extend to higher $n$ harmonics or integral multiple of  $\vec{q}_{1}$, while it is, in contrast, weakly or minimally coupled to the on-axis CDW leaving that incommensurate.
This behavior is in distinct contrast to that of the other two compounds studied, neither of which exhibits higher harmonics of the off-axis modulation vectors.
By way of comparison with Sm$_2$Te$_5$, it is also worth noting that the off-axis CDW peaks in Gd$_2$Te$_5$ lie in a high but flat region, without any sharp peak features, when mapped onto $\chi_{D}(\vec{q})$ of the ditelluride-like double Te layers.
As such, and in contrast to the case of Sm$_2$Te$_5$, the off-axis CDW wave vectors observed in Gd$_2$Te$_5$ appear to get enhanced mainly within the Te single layer by this subtle interaction and develop an extensive commensurate CDW structure.

Although the off-axis lattice modulation is different for Sm$_2$Te$_5$ and Gd$_2$Te$_5$, and although they each exhibit different resonant mechanisms which enhance the off-axis CDW based on interaction with the double or single Te planes respectively, nevertheless, the off-axis CDW for both compounds appears to be principally driven by the single Te planes.
In sharp contrast, Nd$_2$Te$_5$ appears to defy this simple model.
Specifically, the off-axis modulation wave vectors for this compound do not lie close to the global maximum in $\chi_{S}(\vec{q})$ (triangular points in Figure\ \ref{fig:chiq}(c)).
Instead, they are found near \textit{local} maxima which have values about 20$\%$ less than the \textit{global} maximum in both $\chi_{S}(\vec{q})$ and $\chi_{D}(\vec{q})$, although these features are very weak.
Taken at face value, it appears that both single and double planes contribute towards the off-axis CDW in Nd$_2$Te$_5$, although it is impossible from this analysis to determine whether one or the other plays a dominant role.

One of the principle defining features of the off-axis lattice modulation is the huge variation between the three compounds studied, especially given the minimal differences observed in the on-axis wave vector.
Given that the electronic structure is essentially identical for all three compounds, this large variation suggests that differences in the phonon mode characteristics play an important role.
The atomic masses of Nd, Sm and Gd differ by up to 10$\%$, affecting the lattice instability through equation 5.
Presumably, the very well-defined peak feature in $\chi_{D}(\vec{q})$ (Figure\ \ref{fig:chiq}(b)) ensures that the on-axis wave vector remains tied to the wave vector at which this quantity peaks, even as the phonon modes and electron-phonon coupling change.
However, the more poorly defined maximum in $\chi_{S}(\vec{q})$ (Figure\ \ref{fig:chiq}(c)) is apparently not strong enough to completely dominate the electron-phonon coupling to the extent that variation in the phonon characteristics are able to affect the lattice modulation.
This behavior is consistent with that of the single and double layer compounds $R$Te$_2$ and $R$Te$_3$ - the former  having a poorly defined peak in $\chi(\vec{q})$ and a lattice modulation very sensitive to changes in rare earth,\cite{Shin_2005} the latter having a very well-defined peak in $\chi(\vec{q})$, and a lattice modulation that hardly changes across the entire rare earth series.

\section{Conclusion}

In summary, we have presented an alternative method to prepare large, high-quality single crystals of $R_2$Te$_5$ ($R$=Nd, Sm and Gd) via a binary self flux method, and have presented the first evidence for charge density wave formation in this material.
All three compounds exhibit an \textit{on-axis} modulation with $\vec{q}_0\approx0.68$ along the $c^*$-direction, in combination with an \textit{off-axis} superlattice which varies significantly between the three compounds studied. Based on a consideration of contributions to the Lindhard susceptibility from the single and double Te planes of the layered structure, it appears that the on-axis CDW is driven by the double Te planes, whereas the off-axis CDW is principally driven by the single Te planes.
Resonant effects associated with coupling of higher harmonics of these modulation wave vectors to local features in the susceptibility of the double and single Te planes appear to be relevant for Sm$_2$Te$_5$ and Gd$_2$Te$_5$ respectively, stabilizing in the first case mixed commensurate/incommensurate off-axis harmonics, and in the second an extensive commensurate structure associated with just the single Te layers, decoupled from the on-axis modulation.

\section{acknowledgements}

We gratefully thank Robert E. Jones and A. Marshall for technical assistance with EMPA and TEM measurements and analysis.
This work is supported by the DOE, Office of Basic Energy Sciences, under Contract No. DE-AC02-76SF00515. 
Efforts at the Ames Laboratory were supported by the DOE under Contract No. DE-AC02-07CH11358.
IRF is also supported by the Terman Foundation.
Portions of this research were carried out at the Stanford Synchrotron Radiation Laboratory(SSRL), a national user facility operated by Stanford University on behalf of the US Department of Energy, Office of Basic Energy Sciences.

\end{document}